\documentclass[pss]{wiley2sp}
\usepackage{amsmath}
\usepackage{bm}              
\usepackage{w-greek}         

\newlength{\figwidth}
\newlength{\shift}
\newcommand{\fg}[3]
{
\begin{figure*}[ht!]%
\sidecaption
  \includegraphics*[width=\figwidth]{#1}%
  \caption[]{%
#3}
\label{#2}
\end{figure*}}

\begin{document}


\title{Quantum Criticality and Novel Phases:  A panel discussion.}

\titlerunning{QCNP09:  A panel discussion.}

\author{%
Piers Coleman}
\authorrunning{P. ~Coleman}

\mail{e-mail
  \textsf{coleman@physics.rutgers.edu}, Phone:
  +1 732 445 5500 ext 5082}

\institute{
Dept of Physics and Astronomy,\\
Rutgers University,\\
Piscataway,\\
NJ 08854, U.S.A. 
}

\received{XXXX, revised XXXX, accepted XXXX} 
\published{XXXX} 


\abstract{%
%
%
%
\abstcol{
Physicists gathered in august at 
Dresden for a conference about ``Quantum Criticality and Novel
Phases".  As one part of the meeting, nine panelists hosted an open and
free-wheeling discussion  on the topic of the
meeting.  
}
{This article outlines the 
discussions that took place during
at this panel-meeting on the afternoon of August 3rd, 2009.
}}

\maketitle   

\section{Introduction}
 \label{}

Eighty years ago,
physicist Paul Dirac, reflecting on  the new
quantum theory he had played such an important part in developing,
wistfully remarked that
\begin{description}

\item{
{\em ``the underlying physical laws necessary for the
mathematical theory of  ${\dots}$  physics and the whole of
chemistry are thus completely known, and the difficulty is only
that ${\dots}$ these laws lead to equations much too
complicated to be soluble'' }
\hfill{\rm  P. A. M. Dirac $\sim$1929.\cite{dirac} 
}}
\end{description}
Discoveries spanning eight decades have revealed 
Dirac's remark to be one of the great physics 
understatements of all time, for 
understanding the link between the quantum micro-world and our emergent
macroscopic world proves a singular challenge.
Today, we know that the rules of quantum mechanics endow
matter with a propensity to develop unexpectedly simple, yet
completely new kinds of
collective behavior.
%
From the practical perspective of the
material physicist,  emergence means
that the  periodic table is a forge of fabulous
potential, from which wholly new kinds of
material can be crafted, high
temperature superconductors, 
materials with new kinds of multi-functional behavior such as
multiferrocity and 
materials with possibilities that we have yet to discover or even imagine.
But its easy to get lost, and guiding principles are invaluable.

One such principle that has appears increasingly 
fruitful, is to seek materials that lie at the point of instability
between one phase and another: this point is called
a ``quantum phase transition'' (QPT)\cite{Sachdev99}
.
Conventional phase transitions are
driven by thermal motions at finite temperature. 
Quantum phase transitions occur 
at the point where the transition temperature 
is tuned to absolute zero. 
At absolute zero, thermal motion vanishes, yet quantum phase
transitions  are far from static, and can be likened to a melting phenomenon
driven by the zero point quantum motion that arises from
Heisenberg's uncertainty principle. 
Such zero point motion is particularly important at a
second order quantum phase transition, where 
fluctuations in the order
parameter develop an infinite  correlation length and an infinite
correlation time that engulfs the entire material:
such a singular 
point of transition is a ``Quantum Critical Point'' (QCP)\cite{Sondhi97,Sachdev99,Continentino01}.

Experiments show that 
radically new kinds of metallic behavior develop when a quantum
critical metal is warmed to a  finite temperature\cite{review,Loe07.1s}
Such materials 
also have a marked propensity
to nucleate new kinds of order.  The ``dome'' of
superconductivity in the phase diagram of 
high temperature cuprate superconductors is thought by many, to hide a
quantum critical point, but the superconductivity is so robust 
that huge magnetic fields are required to strip away the
superconductivity to reveal the underlying quantum critical point, and
the idea is still controversial. Fortunately, 
similar situations occur in 
heavy fermion materials, such as  $CeRhIn_{5}$, where superconductivity
nucleates around a pressure-tuned
 antiferromagnetic quantum critical point and can
be removed by more modest magnetic field\cite{park06}. 
The important point is that 
quantum criticality appears to provide a vital way
of inducing high temperature superconductivity and other 
novel material behavior; 
it is this potential, together with the dramatic
transformations in metallic behavior that appear to accompany
quantum criticality that motivate the research behind the 
the Dresden ``Quantum Criticality and Novel Phases'' conferences,
discussed in the panel discussion reported here.

Nine panelists Meigan Aronson (Brookhaven
National Laboratory and Stony Brook University, New York, USA),
Piers Coleman (Rutgers University, New Jersey, USA), Philipp Gegenwart
(University of G\" ottingen, Germany), Hilbert von L\" ohneysen
(Karlsruhe University, Germany), Brian Maple (University of
California, San Diego, USA),  Suchitra Sebastian (Cambridge
University, UK), T. Senthil (MIT, USA), Kazuo Ueda (Institute for Solid State
Physics, Tokyo, Japan) and Tomo Uemura (Columbia University, New York, USA).
came together at the conference for a wide-ranging discussion on
this topic. As a prequel to the discussions, each participant posted their
questions and thoughts on a wiki discussion site\cite{wiki}.

In reporting the discussions, I have not followed the original
ordering of speakers, but instead, to group the discussions by topic.
Any mistakes in the rendition of the ideas that were presented at the
discussion are most likely my own, for which I apologize in advance.
I am particularly indebted to those speakers who sent me notes on
their discussion. 

\section{f-electrons as a route to wider understanding}\label{}

One of the persistent threads throughout the discussion, was the
usefulness of f-electron materials as a research basis for studying
quantum criticality and novel phases of strongly correlated materials.
{\sl \bf Meigan Aronson }
emphasized how so much of our understanding about quantum criticality and zero
temperature phase transitions comes from systematic studies of
f-electron based compounds, where the  low energy/temperature range of the
f-bands means that the stability of magnetic order can
be tuned and fully suppressed by modest variations in pressure,
composition and magnetic fields.

{\bf{Kazuo Ueda}} agreed, and described how in Japan, the utility of
f-electron research 
has recently been recognized by the Japanese Ministry of Education through 
the establishment of a distributed research
network, consisting of sevean closely linked research consortia across
Japan working on a wide-band of projects in f-electron physics. 
The network also has funds
for several independent research projects for smaller research groups
some of an applied nature, to link up with the consortium.
One of the vital 
reasons, he said, for a distributed consortium,
is that it makes it possible to share materials , skills, resources and high
quality spectroscopy without them being concentrated at a single institution. 

{\bf Meigan Aronson} referred to the 
mounting evidence from the phase diagrams of systems with
very different microscopic physics - such  as 
organic conductors, and the new Fe-based superconductors - that
unconventional superconductivity may generically occur near magnetic
quantum critical points, suggesting a universality to the overall
behavior which was originally found in f-electron based systems.
 
\section{Strange Metals and Quantum Criticality}\label{}

One of the key  topics, was strange metal behavior and its possible connection
with quantum criticality.
There was a wide ranging  discussion about the meaning of quantum criticality.
Questions were also raised about whether the linkage between bulk
quantum criticality, and the development of strange metal behavior is
indeed always apparent?


{\bf{ Hilbert von Lohneysen}} emphasized that even though
a Quantum Phase Transition (QPT) 
is strictly speaking a zero-temperature instability, its experimental 
manifestations are clearly seen at finite temperatures. In metals
these anomalies give rise to  departures from canonical Landau Fermi
liquid behavior that are often refered to as 
``non-Fermi-liquid'' (NFL) behavior.  In general terms,
he argued, one {\sl can} understand this finite temperature effect of a
quantum critical point in terms of a quantum-classical mapping in
which  temperature sets the maximum time scale for coherent quantum
processes\cite{Cardy96,Sondhi97,Continentino01,Palova08}
, introducing a  finite 
system size $L_{\tau }$ in the time direction.
\[
L_{\tau }= \frac{\hbar }{k_{B}T}.
\]
According to this picture, he emphasized, 
one expects that the “fan-shaped” catchment area of a QCP will extend at
most to temperatures where $\hbar /L_{\tau }$ 
remains small compared with other relevant 
energy scales, such as the Kondo scale $k_{B}T_{K}$.

\begin{figure*}[ht!]%
  \sidecaption
  \includegraphics*[width=.68\textwidth]{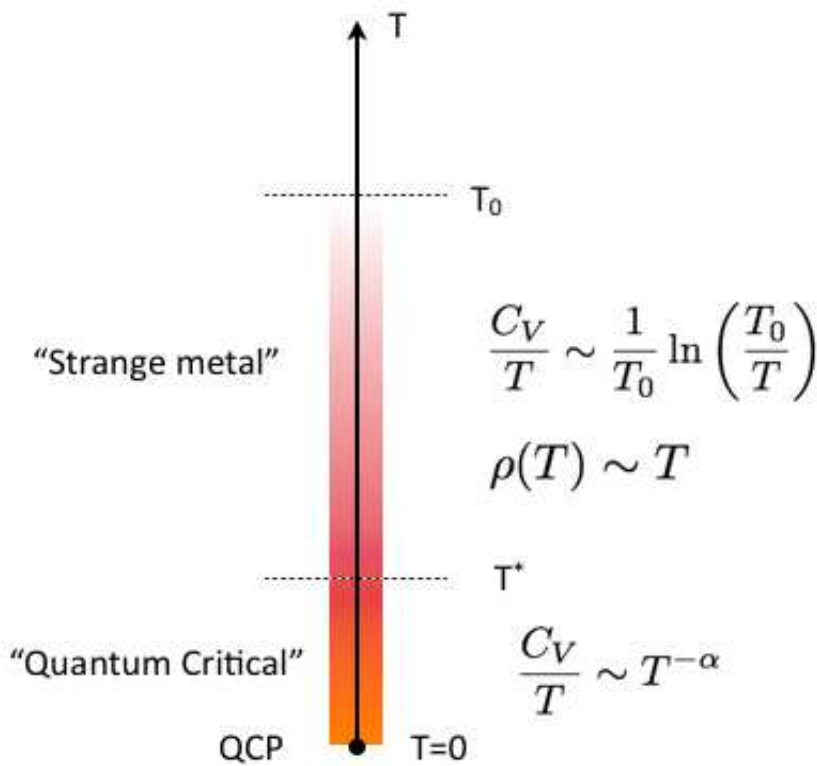}%
  \caption[]{%
The cross-over between ``strange metal'' behavior 
that occurs at high temperatures, to ``quantum critical behavior''
in the vicinity of the quantum critical point.
}
    \label{fig1}
\end{figure*}

Puzzlingly, though, in several
systems, the NFL behavior associated with the QCP extends to
temperatures up to or even in excess of $T_{K}$. 
For instance, in
CeCu$_{5.9}$Au$_{0.1}$ with $T_{K} \approx 5$K, 
the anomalous scaling exponent $\alpha  = 0.75 $
in the dc susceptibility $\chi (T)^{-1}= (A + B T^{\alpha })$extends
up n
to 6 K [1]. This resembles, he remarked,  the
anomalous T-linear resistivity in the cuprate superconductors, 
extending up to $\approx$ 1000K in optimally doped YBa$_{2}$Cu$_{3}$O$_{7}$.
Both 
{\bf{Suchitra Sebastian}}  and
 {\bf{Hilbert von Lohneysen}} questioned
\begin{quote}
{\sl whether the remarkably large temperature ranges over which
strange metal behavior is observed is a signature of the asymptotic
low temperature quantum criticality, or
whether it is some other kind of separate precursor?
}
\end{quote}
For example, in quantum critical
YbRh$_{2}$Si$_{2-x}$Ge$_{x}$
a logarithmic specific heat
$\frac{C}{T}\sim \frac{1}{T_{0}}\ln  (\frac{T_{0}}{T})$ 
is seen over two decades of temperature, where $T_{0}= 24K$, but at
temperatures below $T^{*}\sim 0.3K$, 
the specific heat is seen to cross over to a power-law behavior
$C/T \sim T^{-1/3}$\cite{custers2} (See Fig. \ref{fig1}).

{\bf{Suchitra Sebastian}} asked {\sl what do
unconventional power laws really mean and with what theory should one
compare them?}  It is well know that 
in the approach to a classical critical point, 
that larger temperature deviations $T-T_{c}$ from the critical point are governed by
Ginzburg Landau theory and true critical behavior only develops once
the Ginzburg criterion is violated. Could a cross-over between two
different types of behavior be at work in quantum criticality too?

{\bf{Kazuo Ueda }} remarked that a key observation in f-electron
systems,
is that the mass of electron quasiparticles
becomes very heavy near a quantum critical point, and that
as the characteristic Fermi temperature collapses, strange metal 
behavior is seen to develop. 
In addition to spin fluctuations, he said, there are a variety of
other slow quantum fluctuations that may be important in driving up
quasiparticle mass, including local
orbital fluctuations and anharmonic lattice vibrations. 
Two groups in the new consortium will explore such new mechanisms, he
said. One of the
guiding principles that researchers may follow here, is to look for
systems where the magnetic ion lies at a point of high symmetry.

{\bf{Brian Maple}} expanded further on the need for a 
more general exploration of 
quantum criticality. He argued that the view 
that the identification of strange metal physics with a bulk
quantum critical point may be too restrictive. 
Indeed, key 
anomalous characteristics of strange metal behavior, such as
anomalous temperature dependence in 
\begin{itemize}
\item  resistivity, 
$\rho  (T) \sim \rho_{0}\pm A T^{n}$  ($1\leq n \leq 1.5$ with 
n usually close to 1).

\item specific heat $C(T)/T \sim -lnT,\
T^{-n}$, ($n \sim 0.2
- 0.4$ ) and 

\item magnetic susceptibility 
 $\chi (T) \sim -lnT,\  T^{-n}$ ($n \sim 0.2 - 0.4$), along with 

\item  the observation of $\omega/T$ scaling in neutron scattering
with $\chi ''(q,\omega) \sim f (\omega /T)$. 
\end{itemize}

\noindent are found in widely different circumstance. Not only
only are they found near bulk 
antiferromagnetic quantum critical points, as in YbRh$_{2}$Si$_{2}$, CeRhIn$_{5}$ under
pressure, and CeRh$_{1-x}$Co$_{x}$
In$_{5}$, they are also observed 
at spin-glass quantum critical points in U$_{1-x}$
Y$_{x}$Pd$_{3}$Al$_{2}$,
UCu$_{5-x}$Pd$_{x}$, near ferromagnetic quantum critical points (in
URu$_{2-x}$Re$_{x}$Si$_{2}$ and CePd$_{1-x}$Rh$_{x}$). 
Moreover, he noted, they are also observed far
from any readily identifiable bulk quantum critical point, as in 
(Y$_{1-x}$U$_{x}$Pd$_{3}$
, Sc$_{1-x}$U$_{x}$Pd$_{3}$
 and U$_{1-x}$Th$_{x}
$Pd$_{3}$Al$_{2}$
) prompting a speculation
that there might be a single ion, local quantum-critical
character.  These observations led Brian Maple to ask:\\

{\sl Is
there a more general scenario that encompasses these situations and
presently proposed mechanisms (e.g., Kondo disorder, quadrupolar
Kondo, Griffith’s phase, 2nd order AFM, SG, or FM transition
suppressed to 0 K, etc.)?}\\

\noindent As an example, Maple noted 
Y$_{1-x}$U$_{x}$Pd$_{3}$,
the first f-electron material
in which non-Fermi liquid behavior was observed), 
Sc$_{1-x}$U$_{x}$Pd$_{3}$, and URu$_{2}$Si$_{2}$,
where there is
evidence that the magnetic uranium ion is tetra-valent ($U^{4+}, \ 5f^{2}$),
with two localized f-electrons. In this situation, there is the
possibility of a $\Gamma_3$ or $\Gamma_{5}$ non-Kramers doublet
ground state, setting the stage for a quadrupolar
Kondo effect, originally proposed for Y$_{1-x}$U$_{x}$Pd$_{3}$. 
Perhaps, he proposed,  this scenario,
modified to account for interactions between U ions, can, after all,
account for NFL behavior in these systems.

\section{The mechanism of quantum criticality.}\label{}

The discussion turned to  the mechanism of quantum criticality as
observed in f-electron materials. One of the subjects of particular
interest, concerned the evolution of the Fermi surface through a
quantum critical point. {\bf{Meigan Aronson}} discussed
how measurements have documented
the extent to which the critical fluctuations associated with the
$T=0$ K
transition  affect a wide range of measured quantities, magnetic,
thermal, and transport. Yet at the same time, she noted, 
various experiments such as Hall conductivity,
neutron and de Haas van Alphen suggest the f-electron may
be localizing at a magnetic heavy electron quantum critical point,
rasing her to pose the key question:
\begin{quote}
{\sl how do the anomalous fluctuations at a quantum critical point 
impact the underlying
electronic structure and give rise to the apparent f-electron
delocalization transition that has been found to exist at or near the
quantum critical point in some systems?  } 
\end{quote}

The possibility of a class of quantum phase transitions where the
Fermi surface jumps in area or volume was taken up in detail by
{\bf{Todadri Senthil}}. 
Senthil cited both 
heavy fermion  Quantum Critical Points and Mott Metal-insulator
transitions (which may occur in under and over-doped cuprate
superconductors) as possible examples. He argued that since these 
transitions appear to be second-order, non-Fermi  liquid physics
follows very naturally at such a QCP, but that 
the intuition built up from ``bosonic quantum criticality''
(electrons coupled to a fluctuating order parameter) would  not be relevant.

Senthil described how a model of ``quantum critical Fermi surfaces''
is appropriate to describe the paradoxical combination of a 
first-order jump in the Fermi surface at a second-order phase
transition\cite{senthilfs}.   According to this picture, the large and small
Fermi surface co-exist at a QCP, while the jump $Z$
in momentum-space occupancy is expected to vanish\cite{review,qimiaofs}
at the QCP to be replaced by a Fermi surface with a power-law
singularity in the electron Green functions, similar to what is found
in a one-dimensional Luttinger Liquid (Fig. \ref{fig2}).
\figwidth=0.68 \textwidth
\fg{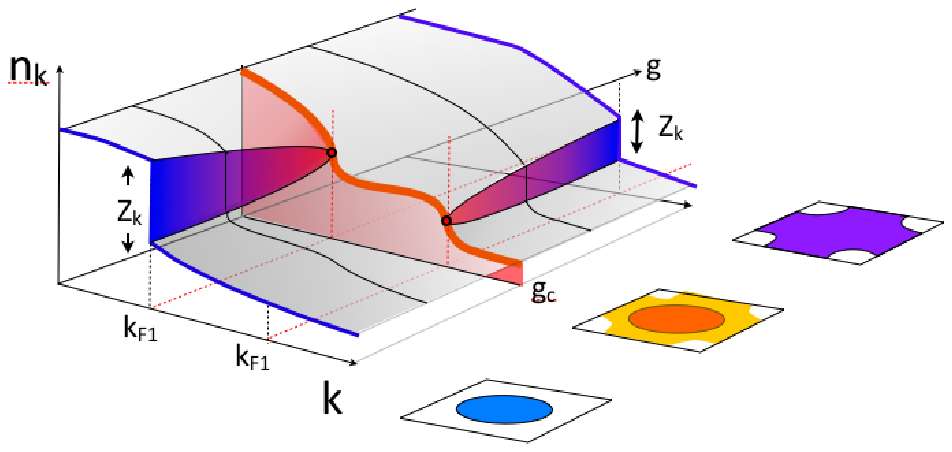}{fig2}{(Color online) Schematic illustrating Senthil's ``quantum
critical Fermi surface'' scenario\cite{senthilfs}, whereby the jump in the occupancy
$Z_{k}$ at the small Fermi surface vanishes at the quantum critical
point\cite{review},
leaving behind a power-law singularity.  As the tuning parameter
passes beyond its critical value $g_{c}$, a new ``large'' Fermi surface develops
at $k=k_{F2}$, describing the heavy Fermi liquid.
}

One of the interesting questions raised by Senthil, is whether
\begin{quote}
{\sl the jump in the Fermi surface volume and the development of
magnetic order are necessarily linked to one-another, or whether are they two
different phenomenon? }
\end{quote}

{\bf{Piers Coleman}} described how \cite{note}
insights from quantum magnetism and recent experiments tend to support
this point of view. 
Imagine he said, connecting a frustrated
antiferromagnet to a conduction sea via a tunable Kondo
interaction.  Various groups\cite{glob1,glob2,glob3,custers} have considered a 
two-dimensional 
phase diagram with $x-$ordinate 
describing the tuning $K= T_{K}/J_{H}$ of the ratio between the 
Kondo temperature and the nearest-neighbor RKKY interaction, and
$y-$ordinate describing the intensity $Q-$ of antiferromagnetic quantum zero-point
fluctuations (which can be tuned for example, by increasing the amount
of frustration). While there 
there is a common  antiferromagnetic phase at small $K$ and $Q$, the 
the paramagnetic ``spin liquid'' at large $Q$ has a small Fermi surface, 
while the the paramagnetic heavy Fermi liquid at large $K$
has a large Fermi, suggesting that the two are separated 
by zero-temperature phase transition (Fig. \ref{fig3}). 
The existence of this transition appears to have been observed in
field-tuning experiments on Ir and Ge doped YbRh$_{2}$Si$_{2}$
(YbRh$_{2-x}$Ir$_{x}$Si$_{2}$\cite{sven} and YbRh$_{2}$Si$_{2-x}$Ge$_{x}$\cite{custers}.
In these materials, there is a field-tuned temperature scale
T$^{*}$(B)  where various anomalies are seen in the Hall constant,
susceptibility and Gr\" uneisen parameter are seen to sharpen up at the
critical field $B_{c}$ where  T$^{*}$(B$_{c}$)$\rightarrow 0$. This
point has been interpreted as the point of field-tuned
transition between a small and a large Fermi surface. 
{\bf{Philipp Gegenwart}} pointed out that in YbRh$_{2}$Si$_{2}$, 
the field-tuned T$^{*}$ scale and the N\' eel temperature 
line and T$_{N}$(B) converge at a single quantum critical point 
but that they separate in Ir or Co or Ge doped
systems\cite{sven,custers}. 
Similar features, though less intensively studied, are seen at higher
magnetic fields in YbGeSi\cite{ybgesi}. 
In both types of material, 
a strange metal phase appears to lie
lie between the ordered antiferromagnet and the heavy electron
state.
{\bf{Philipp Gegenwart }} raised various questions about 
these ideas. He asked, \begin{itemize}

\item  {\sl what is the coincidence of scales in undoped
YbRh$_{2}$Si$_{2}$
 that brings the N\' eel and ``T$^{*}$ line''
together?,
\item What is the nature of the ``spin liquid phase'' that is
predicted to develop in Ir doped  YbRh$_{2}$Si$_{2}$?

\item In YRS, why does pressure fail to influence the 
``T$^{*}$ line''? \cite{tokiwa05}}

\end{itemize}

{\bf Senthil} had several points to make about this kind of phase
diagram, which he epitomized by the phrase {\sl``Quantum is different'' }. These important
differences can occur in many different guises. For example, from
from work on frustrated two-dimensional antiferromagnets,
there are indications that the continuous QPT from antiferromagnet to
spin liquid, or valence bond solid may involve a new kind of
``deconfined criticality'' with emergent fractional degrees of
freedom\cite{deconfined}. He also mentioned the possibility of
topological order\cite{topological1,topological2}
and {\sl ``self organized criticality''} where strange metal phases
containing algebraic order in
space, or time, might develop without fine-tuning to a QCP, such as
a critical spin liquid phase\cite{algebra1,algebra2}.

\fg{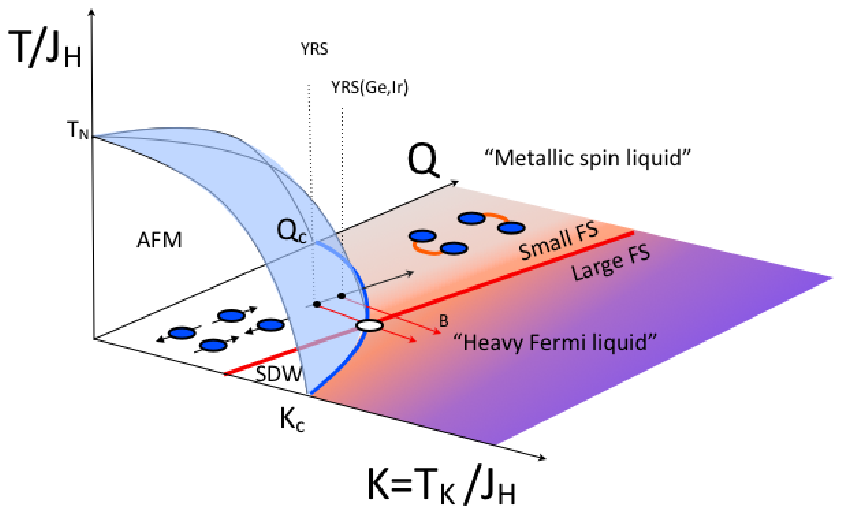}{fig3}{(Color online) Schematic ``global phase diagram'' for heavy
fermion materials, adapted from reference \cite{custers}, showing
``Kondo axis'' tuning the ratio $K=T_{K}/J_{H}$ of the Kondo temperature
to Heisenberg interaction versus the ``Quantum axis'' tuning the
strength of magnetic zero point fluctuations $Q$. Red line denotes
zero temperature quantum phase transition between small and large
Fermi surface states. Labeled are the hypothetical locations of
stoichiometric YbRh$_{2}$Si$_{2}$(YRS) and the same compound, doped
with Ge or Ir (YRS (Ge,Ir)), showing presumed effect of magnetic field
and doping. }

{\bf{ Philipp Gegenwart}}, {\bf{Hilbert von Lohneysen}} and {\bf{Brian
 Maple}} turned the
subject towards the practical, experimental classification and
characterization of different types of quantum critical point.

Gegenwart described how antiferromagnetic quantum
critical points in metals appear to divide into two classes- 
``conventional'', spin-density wave transitions and ``unconventional''
QCP's where the physics appears  more localized.
One of the most useful methods to delineate between different classes of
quantum critical point, he said, is the 
Gr\" uneisen parameter, which measures the ratio 
\[
\Gamma =\frac{V^{-1}dV/dT}{C_{P}}.
\]
Under the assumption that the free energy contains a single
energy scale, so that $F (T)\sim T \phi (T/T^{*} (P))$, 
$\Gamma\propto \frac{d \ln (T^{*})}{dP}$.
It has been shown by scaling arguments\cite{zhu}
that $\Gamma$ diverges in the approach to any QCP. For QCPs of the
SDW type, the critical Gr\" uneisen ratio diverges as $1/T$. By
contrast, a divergence with fractional exponent ($\Gamma \sim
T^{-0.7}$) has been found in Ge-doped YbRh$_{2}$Si$_{2}$
\cite{kuchler03}, which appears to be compatible with the predictions
of a locally quantum critical senario.  If $\Gamma$ has no divergence,
one can exclude a 
generic QCP as the origin of non-Fermi liquid behavior.
%
%
Gegenwart noted that experimentally, both types of QCP (SDW and local) are
observed in different heavy fermion metals. Unconventional transitions pose the greatest challenge, with
many open questions:
\begin{itemize}
\item Which types of unconventional exist?
\item What are the conditions they arise from?
\item Which observed features are generic and which are material specific? 
\end{itemize}

Gegenwart turned to discuss the field-tuned quantum criticality of 
of YbRh$_{2}$Si$_{2}$, one 
where the magnetic Gr\" uneisen parameter $\Gamma_{M}=
-\frac{dM/dT}{C_{H}}$
diverges as $\Gamma_{M} = G_{r}/ (H-H_{c})$ with $G_{r}=-0.3$\cite{tokiwa09}.
Gegenwart asked whether this result might be 
consistent with the critical Fermi surface model of Senthil?

{\bf{Hilbert von Lohneysen}} 
emphasized that the difficulty with any kind of global phase diagram,
is that we do not understand the roles of different 
tuning parameters at a QPT. He pointed out that 
there are only few systems where different parameters
have been employed to tune a QCP. In the case of CeCu$_{6-x}$Au$_{x}$, three
different tuning parameters have been employed: chemical doping with
gold, concentration x, hydrostatic pressure $P$, and magnetic field $B$. While P and x tuning
lead to the same T dependencies in resistivity $\rho $ and specific
heat $C$, both suggestive of a local QCP, field tuning shows dependencies in $\rho $ and C
that are more indicative of a spin-density wave (SDW)-type QCP [2]. 
The different scaling behavior for x
and B tuning has been corroborated by inelastic neutron scattering
experiments, directly measuring the critical fluctuations [1,
3]. These experiments put definite constraints on the construction of
a multi-parameter zero-temperature (“global”) phase diagram. More
experiments along theses lines are certainly needed.

{\bf{ Brian Maple}} discussed how little we understand about ferromagnetic
quantum critical points in $f-$electron systems.  Unfortunately, he said,
most ferromagnetic systems exhibit a first order QPT under pressure,
but a second order QCP under chemical doping. 
He introduced
URu$_{2-x}$Re$_{x}$Si$_{2}$ as a fascinating new development in this respect.
Undoped URu$_{2}$Si$_{2}$ exhibits ``hidden order'', 
but on doping 
with rhenium, he said, the transition temperature of the hidden order
is continuously suppressed to zero at $x_{c}\approx 0.15$.
The QPT where the
hidden order disappears coincides with a sudden appearance of
ferromagnetism, with a Curie temperature $T_{c}$ that grows linearly
with $x-x_{c}$. This ferromagnetic phase exhibits non-Fermi liquid
properties: a logarithmic temperature dependence of the specific heat
and an anomalous temperature dependence of the resistivity.
The logarithmic specific heat  exhibits a remarkable parabolic
dependence on doping,
\[
\gamma (x,T) \sim -(x-x_{c}) (x_{2}-x) \times \frac{1}{T_{0}}\ln T,
\]
where $x=x_{2}=0.6$, even though ferromagnetism continues to higher doping.
Maple reported that 
a careful study of evolution of the
temperature exponent of the susceptibility $\chi \sim t^{-\gamma}$,
the field and temperature 
dependence of the magnetization $M\sim H^{1/\delta }, \ t^{\beta }$
(linked by the relation $\delta-1 = \gamma/\beta $) shows that as a function of doping, $\gamma$ and $\delta $ grow linearly with $x-x_{c}$
while $\beta \approx 1$ is constant\cite{maplefm}.  This fascinating physics awaits
a theory.

{\bf{Tomo Uemura}} and {\bf{Suchitra Sebastian}} 
both returned the discussion of the nature of the soft quantum modes
near a QPT. Both 
raised the issue of the relationship of spin and charge degrees of
freedom in materials close to a quantum critical point, and indeed, 
whether spin is the most important slow degree of freedom. 


{\bf{ Tomo Uemura}}  raised two interesting points in connection with
the soft-modes at a QCP. First, he asked, what is the role of 
importance of first-order quantum phase transitions.  Even though many quantum
phase transitions are first order, they may still exhibit a variety of
important soft modes that while strictly speaking, remain gapped at
the transition, still influence the physics over a wide finite
temperature range. Such soft modes  are the analog of ``roton modes''
in He$^{4}$, but they might occur in the spin, the charge and even the
phase channel. In general, even if the QCP is first order, 
the energy of such soft modes can be used as an indicator of the
closeness to a competing state\cite{tomo1,tomo2}. 

Uemura speculatively introduced the idea of 
``resonant spin-charge coupling'' -  while most of us think of the
charge and spin modes of strongly correlated systems as separate
degrees of freedom,
\begin{quote}
{\sl  is it possible, he mused, that 
that slow spin and charge modes become resonantly coupled? }
\end{quote}
Uemura 
he showed how the superconducting transition temperature $T_{c}$ in a
wide range of unconventional superconductors scales with the Fermi temperature
$T_{F}$ (obtained from the linear specific heat) and the
spin-fluctuation scale (obtained from the magnetic susceptibility), namely
\[
T_{c}\propto T_{F},\ T_{SF}.
\]
Such scaling relationships seem to hold over many decades of variation
in $T_{c}$.  Conventionally, these kinds of scaling relationships are
interpreted in terms of an anisotropic pairing within a Fermi liquid,
in which the a single renormalized Fermi temperature of the pairing
electrons also governs the characteristic spin fluctuation scale.
Uemura argues that an alternative way to interpret tracking between
$T_{F}$ and $T_{SF}$ illustrates a resonance between spin and charge
fluctuation modes.

{\bf Suchitra Sebastian} viewed the problem from another perspective. 
She asked:
\begin{quote}
{\sl should we think of a 
separation of charge and spin Quantum critical points?}
\end{quote}
Such a separation has, she pointed out, been used to understand the
effect of pressure in CeCu$_{2}$Si$_{2-x}$Ge$_{x}$, where the
superconducting transition temperature is seen to exhibit two separate
maxima as a function of pressure - a lower pressure maximum 
in the vicinity of a magnetic plus a higher pressure maximum 
in the vicinity of a valence instability of the
material\cite{cecu2si2ge,miyake}. But could this, she asked, be part
of a much more general phenomenon? Could one, for example, understand
the two transitions seen in Ge and Ir doped YbRh$_{2}$Si$_{2}$ as a
spin and a charge critical point?\cite{sven,custers}. Sebastian also showed 
de Haas van Alphen measurements on  CeIn$_{3}$, in
where a field-induced 
quantum phase transition is observed within the antiferromagnetic
phase at which the effective mass of the quasiparticles diverges while
the orbit areas collapse to zero\cite{suchitra1}.  
A similar divergence in quasiparticle effective masses has also been
recently seen in the de Haas van Alphen measurements on
under-doped YBa$_{2}$Cu$_{3}$O$_{6+x}$\cite{suchitra2}.
Could these, she asked, be further examples of this general phenomenon - QPT involving charge?


\section{New Phases, superconductivity and the d-f connection}\label{}

Throughout the discussion, physicists constantly returned to the
theme of superconductivity, its connection with quantum criticality
and the usefulness of f-electron research as a platform for
understanding ordering phenomenon at higher energy scales, d-electron
transition metals. The ``d-f'' connection  was discussed by several of
the panelists.

{\bf{ Tomo Uemura}} discussed how the broad trends in $T_{c}$ with
superfluid density which cut across anomalous f- and d-electron
superconductors support the idea that we should endeavor to understand
these phenomenon within a unified framework.  As part of this
framework, he argued, there are two extreme limits to consider - one
extreme - that of BCS pairing, while on the other - that of
Bose-Einstein condensation of pre-formed pairs.  Uemura argues that
the scaling of $T_{c}$ with superfluid density is an indication that
anomalous superconductivity lies at the BCS-BEC cross-over 
between these two extreme regimes.

{\bf{Kazuo Ueda}} 
Discussed how a major part of the new 
Japanese f-electron consortium, is the discovery of new materials,
with new types of broken symmetry ground-state
will broaden our understanding of strong correlation, helping
the d-f connection.
This thrust has two components, he said - \begin{itemize}
\item By extending the search for novel f-electron behavior along the
rare earth series - at one end of the series, from Cerium  to Paladium
compounds, and at the other end of the series, from Ytterbium to
Thullium (Tm) materials.

\item By pushing up to higher energy scales, intermediate between the
4f- and 3d materials through the exploration of 5f correlated electron
materials. By going from the Cerium 115 materials to related
transuranic materials, PuCoGa$_{5}$ and NpAl$_{2}$Pd$_{5}$, it has proved
possible to substantially raise $T_{c}$. Are there other examples of
this trend?
\end{itemize}

{\bf{Meigan Aronson}} discussed the difficulties in making the d-f
connection, noting that 
while an 
extensive body of measurements on itinerant ferromagnets
such as ZrZn$_{2}$ and MnSi, led the way in establishing magnetic
quantum criticality in a d-electron context, 
\cite{grosche95,pfleiderer01},
unlike their f- counterparts, 
these systems
have `large' Fermi surfaces in both the ordered and paramagnetic
regimes, where the d-electrons are included in the Fermi surface
\cite{yates03}. While 
systems such as V$_{2}$O$_{3}$ \cite{maiti01} and
Ni(S$_{1-x}$Se$_{x}$)$_{2}$ \cite{matsura98} 
are considered exemplars of Mott-Hubbard
physics, with a first order 
transition from strongly correlated metal to localized moment
insulator 
\cite{kotliar04}, she points out we still have not discovered
a transition metal compound that is the analog
of magnetic field tuned 
YbRh$_{2}$Si$_{2}$ \cite{gegenwart07} or pressurized
CeRhIn$_{5}$\cite{park06}, 
where quantum critical points are the source of both strong
quantum critical fluctuations and transitions where an f-electron is
delocalized, driving the Fermi surface from small to
large\cite{shimuzu}. 

\section{Conclusion}\label{}
\figwidth=0.6\textwidth
\fg{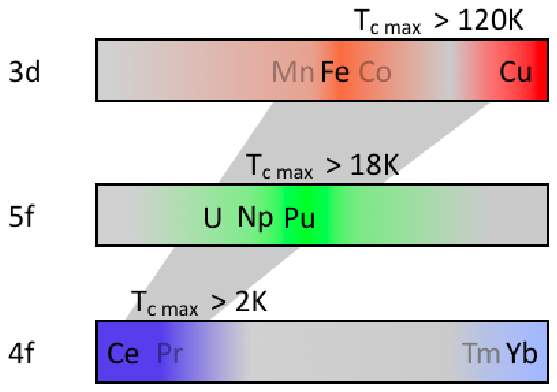}{fig4}{(Color online). The ``d-f'' connection.  As one moves from 4f to 5f
through 3d compounds, the characteristic scale of the d/f electrons
rises, and with it the maximum observed superconducting
temperature. Shaded elements in each series denote the d/f metals whose
compounds are give the currently highest superconducting $T_{c}$.}

The panel discussion prompted lively debate throughout the QCNP09
meeting, and appears to provide a useful model for future scientific
conferences of this time.    The study of quantum criticality and its
intimate relationship with material physics 
and the emergent quantum mechanics of the periodic table, make it an
area of burgeoning discovery.  As a reporter on this event, I'd like
end with a quote from Meigan Aronson at this event, on the prospects
of a future d-f connection surrounding quantum criticality and high
temperature superconductivity (see Fig. \ref{fig4}):

\begin{quote}
{\sl ``So far, our interest has focused on ferromagnetic
systems, which are ultimately discontinuous, and not quantum critical,
when the Curie temperature becomes sufficiently low 
\cite{belitz05,uemura07}.  
Whether a similar behavior will be found in (d-electron) quantum
critical antiferromagnets remains an interesting and controversial
question, little explored due to a lack of suitable host
systems. Still, these are the compounds in which unconventional
superconductivity is presumed to be most likely, particularly if it is
possible to realize the strongly correlated state found on the
metallic side of a Mott-Hubbard transition.  
For these reasons, it
seems more pressing than ever to refresh our interest in intermetallic
compounds where the magnetic entity derives from transition metal
moments.'' }
\end{quote}

\begin{acknowledgement}
I am indebted to all the participants in the QCNP09 panel discussion
for providing their view-graphs in advance of the session, and subsequently
notes on their presentations.  Piers Coleman is supported by the Department of Energy grant DE-FG02-99ER45790.
\end{acknowledgement}

%
%

\providecommand{\WileyBibTextsc}{}
\let\textsc\WileyBibTextsc
\providecommand{\othercit}{}
\providecommand{\jr}[1]{#1}
\providecommand{\etal}{~et~al.}

\end{document}